\documentstyle[floats,aps,epsfig,prl]{revtex}

\begin{document}
\draft

\twocolumn[\hsize\textwidth\columnwidth\hsize\csname
@twocolumnfalse\endcsname

\title{Exponential decay properties of Wannier functions
and related quantities}

\author{Lixin He and David Vanderbilt}
\address{Department of Physics and Astronomy, Rutgers University,
Piscataway, New Jersey 08855-0849}

\date{January 19, 2001}
\maketitle

\begin{abstract}
The spatial decay properties of Wannier functions and related
quantities have been investigated using analytical and numerical
methods.  We find that the form of the decay is a power law times an
exponential, with a particular power-law exponent that is universal
for each kind of quantity.  In one dimension we
find an exponent of $-$3/4 for Wannier functions, $-$1/2 for the
density matrix and for energy matrix elements, and $-$1/2 or $-$3/2
for different constructions of non-orthonormal Wannier-like
functions.
\end{abstract}

\vskip1pc]



\narrowtext

\baselineskip=11.5pt plus 2pt minus 1pt

A growing interest in localized real-space descriptions of the
electronic structure of solids has been motivated by the
development of computationally efficient ``linear-scaling''
algorithms\cite{galli-on,goedecker-on} and by the desirability
of a local real-space mapping of chemical
\cite{marz-wann,silv-wann} and dielectric \cite{souza-h2,water-wann}
properties.  A primary avenue to such a description is the use of
Wannier functions\cite{wannier,kohn59,blount} (WFs), i.e., a set of
localized wavefunctions $w_{\bf R}({\bf r})$ obtained from the Bloch
functions $\psi_{\bf k}({\bf r})$ by a Fourier-like unitary
transformation.  A closely related approach is to represent the
electronic structure in terms of the density matrix
$n({\bf r},{\bf r'})$.  It is thus not surprising to find
considerable recent interest in the localization properties of the
WFs\cite{marz-wann} and of the density matrix\cite{goedecker-dm,arias-dm}.

In a classic 1959 paper, Kohn proved, for the case of a
centrosymmetric crystal in one dimension (1D), that the WFs have an
``exponential decay'' $w(x)\approx e^{-hx}$, where
$h$ is the distance of a branch point from the
real axis in the complex-$k$ plane\cite{kohn59}.
More precisely,
\begin{equation}
\lim_{x \rightarrow \infty } w(x)\,e^{qx}=\left \{ \begin{array}{ll}
0,\quad & \mbox{$q<h$} \\ \infty,  & \mbox {$q>h$}\;\;.  \end{array} \right.
\label{eq:kohn}
\end{equation}
The density matrix has a similar decay
$n(x,x')\approx e^{-h|x-x'|}$.  The exponential decay of the WFs has
since been proven for the general 1D \cite{cloizeauxa} and single-band 3D
\cite{nenciu} cases, and that of the density matrix (more precisely,
of the band projection operator) has been proven in general
\cite{cloizeauxa}.  The energy matrix elements
$E(R)=\langle w_R\vert H\vert w_0\rangle$, with $w_R(x)=w(x-R)$
and $R=la$ a lattice vector, are also expected to have a similar
decay, $E(R)\sim e^{-hR}$.

The purpose of this Letter is to address two questions.  First,
Eq.~(\ref{eq:kohn}) allows considerable freedom; in fact, it is
consistent with
\begin{equation}
w(x)\approx x^{-\alpha}e^{-hx}
\label{eq:powexp}
\end{equation}
for {\it any exponent} $\alpha$, i.e., a decay which could be
faster ($\alpha>0$) or slower ($\alpha<0$) than pure
exponential.  Does such a power-law prefactor exist, and if so,
what is the exponent $\alpha$?  Second, it has long been
understood that relaxation of the orthogonality constraint
$\langle w_0\vert w_R\rangle=\delta_{0,R}$ can give ``more
localized'' Wannier-like functions \cite{anderson,bullett,gallipar}.
In what sense are these more localized -- a larger $h$,
or a larger $\alpha$ for the same $h$, or only a smaller prefactor
of the tail?
We show that the power-law prefactors of Eq.~(\ref{eq:powexp})
{\it do} exist, and that the various quantities have a common
inverse decay length $h$ but different exponents $\alpha$.
In 1D we find that $\alpha=3/4$ for usual (orthonormal)
WFs, $\alpha=1/2$ for $n(x,x')$ and $E(R)$, and $\alpha=1/2$
or $\alpha=3/2$ for two different constructions of non-orthonormal
Wannier-like functions (NWFs).  The NWFs of superior decay
($\sim x^{-3/2}e^{-hx}$) can be constructed by a projection method
as duals to a set of trial functions.  These results may have
important implications for the design and implementation of
efficient real-space electronic-structure algorithms.

We first review the central results of the pioneering work of Kohn
\cite{kohn59}, who considers a centrosymmetric potential
of period $a$ in 1D.  The WFs are constructed as
\begin{equation}
w_n(x-R)=w_{nR}(x)=\frac{a}{2\pi} \int_{-\pi/a}^{\pi/a} e^{-ikR}\psi_{nk}(x) \, dk
\label{eq:wannier}
\end{equation}
with the phases of the Bloch functions $\psi_{nk}$ chosen
as in Sec.~6 of Ref.~\cite{kohn59}.  The exponential decay of the
WFs is then governed by the positions of branch points in the
``complex band structure'' $E_n(k)$ constructed by regarding
complex $E_n$ to be a function of complex $k$ via analytic
continuation from the real axis \cite{kohn59,blount}.  Specifically,
there is a Riemann
sheet $E_n(k)$ for each band $n$, and the branch points $k_n$ are the points at
which the sheets are connected, $E_n(k_n)=E_{n+1}(k_n)$.  These
are located at
\begin{equation}
k_n=\left \{ \begin{array}{ll}
\pi /a \pm ih_n, \quad & \mbox{$n$ even} \\
\pm ih_n, & \mbox{$n$ odd} \end{array} \right.
\label{eq:branch1}
\end{equation} 
and at translational image locations $k_n^{(m)}=k_n+2\pi m/a$ for
integer $m$.  $E_n(k)$ and $\psi_n(k)$ are thus analytic functions
in the strip $|{\rm Im}(k)|<\bar{h}_n$ where
$\bar{h}_n$=$\min (h_{n-1},h_{n})$.
Kohn's main result \cite{kohn59} is that the decay of the WF for the $n$'th
band is as $w_n(x)\approx e^{-\bar{h}_n|x|}$ in the sense of
Eq.~(\ref{eq:kohn}).  In what follows we
restrict our attention to the bottom band ($n$=0), for which
$\bar{h}_0=h_0$ (henceforth just $h$).  The relevant
branch point in the upper half-plane is $k_0=\pi/a+ih$ and the
expected Wannier decay is $w(x)\approx e^{-hx}$.

To confirm this decay numerically, we first
choose a simple 1D model Hamiltonian having a periodic potential
$U(x)=\sum_m V_{\rm at}(x-ma)$ constructed as a sum of Gaussian
``atomic'' potentials $V_{\rm at}(x)=(V_0/b \sqrt{\pi})
e^{-x^2/b^2}$.  Here $a$ is the lattice constant and $V_0$ and
$b$ control the depth and width of $V_{\rm at}$.
We choose units such that $m$=$\hbar$=$e$=1 and keep
$a$=1 and $V_0$=$-$10 fixed while adjusting $b$ to vary the gap.
The Bloch functions are computed on a mesh of 200 $k$ points by
expanding in 401 plane waves
and the WF at $R=0$ is then constructed according to
Eq.~(\ref{eq:wannier}) using 128-bit arithmetic.

\begin{figure}
\epsfxsize=2.5in
\centerline{\epsfbox{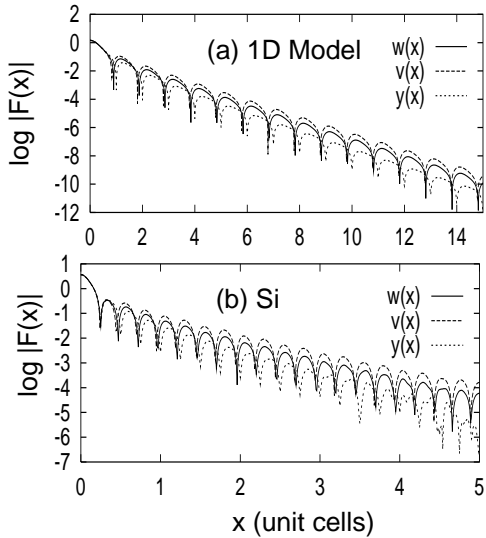}}
\vskip 0.5cm
\caption{Decay of normalized WFs $w(x)$ and NWFs $v(x)$ and $y(x)$.
(a) 1D model (see text) with $b$=0.3 and $V_0$=$-$10.
(b) 3D Si plotted along the [110] direction.}
\label{fig:lwannier}
\end{figure}

\begin{figure}
\epsfxsize=2.9in
\centerline{\epsfbox{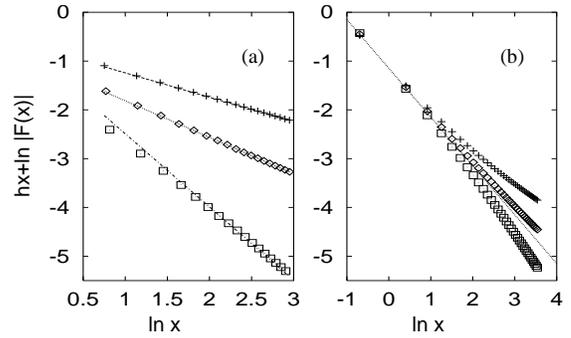}}
\vskip 0.5cm 
\caption{(a) As in Fig.~\protect\ref{fig:lwannier}(a) but
plotted so that slope reveals exponent $-\alpha$
of Eq.~(\protect\ref{eq:powexp}).
(b) Same for $b$=0.6 (nearly-free electron case) showing crossover.
Pluses, diamonds, and squares represent $v$, $w$, and $y$,
respectively.}
\label{fig:power}
\end{figure}

The resulting decay of the WF for $b$=0.3 is shown as the solid
line in Fig.~\ref{fig:lwannier}(a).
In this semilog plot, the approximate linearity of the
peaks is consistent with the expected exponential decay, but
there is a slight curvature that can be
analyzed further.  To do so, we first computed the
$E_n(k)$ along $\pi/a +i\kappa$ for real $\kappa$ and
defined $h$ to be the value of $\kappa$ at which $E_0=E_1$.
For $b$=0.3 we find $h$=1.28869.
In Fig.~\ref{fig:power}(a) we then plot (diamonds)
$hx+\ln|w(x)|$ vs.~$\ln(x)$ for each peak of $\ln|w(x)|$.
A pure exponential decay $w(x)\approx e^{-hx}$
should yield a horizontal line in such a plot; instead, the data
appears linear with a slope of $-3/4$, indicating that
\begin{equation}
w(x) \approx x^{-3/4}\,e^{-hx} \;\;.
\label{eq:wdecay}
\end{equation}
A similar plot (not shown) for
\begin{equation}
E(R)=\langle w_R\vert H\vert w_0\rangle= \frac{a}{2\pi}\int dk\, e^{ikR}E(k)
\;\;.
\label{eq:er}
\end{equation}
suggests that
$E(R)$ shares the same inverse decay length $h$ but has a different
power-law exponent,
\begin{equation}
E(R) \approx R^{-3/2}\,e^{-hR} \;\;.
\label{eq:Rdecay}
\end{equation}
Naturally $h$ changes if the potential parameter $b$ is varied, but
we find that the power-law exponents of $-3/4$ and $-3/2$ do not.
It thus appears that these exponents are a universal feature of
electron bandstructures in 1D.

In order to gain an analytic understanding of this behavior,
we consider first the simpler case of the energy-band Fourier
transform $E(R)\leftrightarrow E(k)$. Kohn showed that the
expansion of $E(k)$ about $k_0=\pi/a+ih$ takes the form \cite{kohn59}
\begin{equation}
E(k)=E_0+\gamma\,(k-k_0)^{1/2}+\cdots
\label{eq:ebranch}
\end{equation}   
with higher terms of order $(k-k_0)^1$, $(k-k_0)^{3/2}$, etc.  The form
of this expansion arises from the requirement that $E(k)$ come back
to itself if $k$ traverses a closed path winding twice around $k_0$,
consistent with the picture of two Riemann sheets touching at $k_0$.

\begin{figure}
\epsfxsize=2.2in
\centerline{\epsfbox{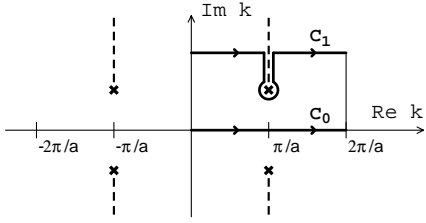}}
\vskip 0.5cm 
\caption{Branch points ($\times$), cuts (dashed lines), and
integration contours ($C_0$ and $C_1$) in the complex-$k$
plane.}
\label{fig:contour}
\end{figure}

Now there are well-known mathematical results that relate the behavior
of a function near a branch point to the asymptotic decay of its Fourier
transform \cite{math}.
The following lemma is useful here.  Let $f(k)$ be a
periodic function $f(k)=f(k+2\pi/a)$ that has a leading behavior
\begin{equation}
f(k)=f_0+\gamma\,[i(k-k_0)]^\beta
\label{eq:fdef}
\end{equation}
when expanded at the branch point $k_0=\pi/a+ih$.
Its Fourier series coefficients are given by
\begin{equation}
F(x) = \int_{C_0} f(k)\,e^{ikx}\,dk
\label{eq:lemma2}
\end{equation}
at $x=ma$ for integer $m$.  As shown in Fig.~\ref{fig:contour}, the
contour $C_0$ initially lies along the real axis.
However, $f(k)\,e^{ikx}$ is invariant
under $k\rightarrow k+2\pi/a$, and assuming that no other branch points
or poles intervene, the contour can be deformed to become $C_1$
as shown in Fig.~\ref{fig:contour}.  The exponential
smallness of $e^{ikx}$ for large $x$ kills the integrals along the horizontal
segments, and the dominant contribution
to the $C_1$ integral comes from the vicinity
of $k_0$.  Using the contour-integral definition of the
Gamma function\cite{gamma},
\begin{equation}
|F(x)| \simeq \gamma\,B(\beta)\,x^{-(1+\beta)}\,e^{-hx} \;\;,
\label{eq:lemma3}
\end{equation}
where $B_\beta= 2\sin(\beta \pi)\,\Gamma(1+\beta)$.

Eq.~(\ref{eq:lemma3}) now allows us to understand the observed
behavior of quantities such as $E(R)$ and $w(x)$.  For example,
since $E(k)$ in Eq.~(\ref{eq:ebranch}) has $\beta=1/2$, we confirm
that $E(R)\approx R^{-\alpha}\,e^{-hR}$ with $\alpha=1+\beta=3/2$.
Similarly, to understand the decay of $w(x)$, we
need to know the behavior of $\psi_k(x)$ regarded as a function
of $k$ near the branch point $k_0$.  Once again
Kohn \cite{kohn59} provides the needed result
$\psi_k\approx(k-k_0)^{-1/4}$.
Sure enough, $\beta=-1/4$ gives a decay
$w(x+R)\approx R^{-3/4}e^{-hR}$ for small $x$ and $R\gg a$.  In
other words, $w(x)\approx x^{-3/4}e^{-hx}$ for large $x$,
as obtained numerically from Fig.~\ref{fig:power}(a).

We can summarize the information about both the $k$- and
$x$-dependence of $\psi_k(x)$ near the branch point as
\begin{equation}
\psi_k(x) = A_0(x)\,q^{-1/4}+A_1(x)\,q^{1/4}+ \cdots,
\label{eq:psi}
\end{equation}
where $q=i(k-k_0)$.
(All such terms have powers that are odd-integer
multiples of 1/4, consistent with $\psi_k(x)\rightarrow-\psi_k(x)$
when traversing a closed path winding twice around $k_0$.)
$A_0(x)$ and $A_1(x)$ are real functions obeying
$A_n(x+a)=e^{ik_0a}\,A_n(x)$.

The locality of the density matrix (i.e., the band projection
operator) is also very important.  For example,
many linear-scaling algorithms are based on a direct solution
for the density matrix \cite{galli-on,goedecker-on,lnv}.
We can write
\begin{equation}
n(x',x)=\frac{a}{2\pi}\int_{C_0} \psi_{-k}(x') \, \psi_k(x) \, dk
\label{eq:density}
\end{equation} 
where, following Kohn \cite{kohn59}, we have substituted $\psi_k^*(x')$
by $\psi_{-k}(x')$  in order that the integrand of Eq.~(\ref{eq:density})
should remain analytic off the real axis.  The behavior of
$\psi_{-k}(x)$ near the branch point is $\psi_{-k}(x)
\approx A_0(-x)\,[i(k-k_0)]^{-1/4}$.
The integrand of Eq.~(\ref{eq:density}) then takes the
form $\psi_{-k}(x') \psi_k(x) \approx A_0(-x')A_0(x)[i(k-k_0)]^{-1/2}$.
Applying Eq.~(\ref{eq:lemma3}) yields $n(0,x)\approx x^{-1/2}\,e^{-hx}$
for large $x$, and more generally,
$n(x',x)\approx(x-x')^{-1/2}\,e^{-h(x-x')}$ for $x \gg x'$.
This $\alpha=1/2$ behavior of the decay has been confirmed from
numerical plots (not shown) similar to Fig.~\ref{fig:power}(a).

We have so far shown that $E(x)$,
$w(x)$, and $n(0,x)$ all have a decay of
the form $x^{-\alpha}\,e^{-hx}$ with a common $h$ but with different
(universal) exponents $\alpha_E$=3/2, $\alpha_w$=3/4, and
$\alpha_n$=1/2.
The energy matrix elements thus have the fastest decay, and
the density matrix the slowest.

One may next ask whether it is possible to find non-orthonormal Wannier-like
functions (NWFs) with a faster decay than those of the orthonormal WFs $w(x)$
\cite{anderson,bullett,gallipar}.
We explore
this question in the context of band-projection methods
\cite{cloizeauxb,goedecker-proj,stephan98}.  We find
that a naive application of the projection technique actually
generates NWFs with a {\it slower} decay, while a modified ``dual
construction'' approach {\it does} give improvement as measured by
the exponent $\alpha$.

The basic idea of the projection technique is to start with a trial
function $t(x)$ and generate a Wannier-like function $v(x)$ by
acting with the band-projection operator
$\hat{P}=\sum_k |\psi_k\rangle \langle \psi_k |$,
%
i.e., $|v_R\rangle$=$\hat{P}\,|t_R\rangle$.
%
Here $|t_R\rangle$ corresponds to the translational image
$t(x-R)$ of $t(x)$ in cell $R=na$, and similarly for $|v_R\rangle$.
The trial functions can be Gaussian functions, atomic or molecular
orbitals, etc.  The $|v_R\rangle$ are NWFs having overlap
%
$S_{0R}$ = $\langle v_0|v_R\rangle = \langle t_0|\,\hat{P}\,|t_R\rangle$.
%
Numerical investigations on C and Si by Stephan and Drabold indicated
that the projected functions $v(x)$ are {\it not} more localized than the
true WFs $w(x)$ \cite{stephan98}.  This should not be
surprising; introduction of NWFs may give flexibility to generate
more localized orbitals, but this flexibility needs to be used to
advantage.  To do so, we introduce {\it dual} functions $y(x)$
defined via
%
$|y_0\rangle$=$\sum_R (S^{-1})_{0R} \, |v_R\rangle$,
%
so that $\langle y_0 |v_R\rangle=\delta_{0R}$ and also
$\langle y_0 |t_R\rangle=\delta_{0R}$.  This latter equation means
that $y(x)$ is orthogonal to the trial function at every site except
$R$=0, suggesting that $y(x)$ may be especially well localized.

Numerical tests of the decay of (normalized versions of) $v(x)$
and $y(x)$ are shown as dashed and dotted curves respectively in
Figs.~\ref{fig:lwannier}(a) and \ref{fig:power}(a).
The trail function used is a $\delta$-function on the atomic
site, but use of other narrow trial functions gives similar results.
It clearly appears that $\alpha=1/2$ and $3/2$ for $v(x)$ and $y(x)$
respectively, to be compared with $\alpha=3/4$ for $w(x)$.
Thus, the simple projected functions $v(x)$ actually have a {\it
slower} decay that the WFs $w(x)$, but the duals $y(x)$ have a
much faster decay than either of them.

These results can be explained by the complex analysis of
the Bloch-like functions $v_k(x)$ and $y_k(x)$ that are related to
$v(x)$ and $y(x)$ in the same way that $\psi_k(x)$ is related to
$w(x)$.  Defining
\begin{equation}
\eta(k)=\int_{-\infty}^{\infty} \psi_{-k}(x)\,t(x)\,dx \;\;,
\end{equation}
it follows from $|v_k\rangle=|\psi_k\rangle\langle\psi_k|t\rangle$
that $|v_k\rangle=\eta_k\,|\psi_k\rangle$.  Also the Fourier transform
of $S(0,R)$ can be seen to be $S(k)=\eta^2(k)$,
so that $|y_k\rangle=|\psi_k\rangle/\eta(k)$.  In the vicinity of
$k_0$ we have
\begin{equation}
\eta(k) = \eta_0\,[i(k-k_0)]^{-1/4}+\eta_1\,[i(k-k_0)]^{1/4}+\cdots
\label{eq:eta1}
\end{equation}
where $\eta_n = \int_{-\infty}^{\infty}	 A_n(-x) \,t(x)\,dx$.
Moreover,
\begin{eqnarray}
v_k(x) &=& \eta_0\,A_0(x)\,q^{-1/2}+ \cdots \;\;, \nonumber \\
w_k(x) &=& A_0(x)\,q^{-1/4}+ \cdots \nonumber \;\;, \\
y_k(x) &=& {1\over\eta_0}\,\left\{\,A_0(x)+\widetilde{A}_1(x)
\, q^{1/2}+\cdots \,\right\} \;\;,
\label{eq:yk}
\end{eqnarray}
where $q=i(k-k_0)$ and $\widetilde{A}_1(x)= A_1(x)-(\eta_1/\eta_0)\,A_0(x)$.
The leading term in $y_k(x)$ gives no singularity,
so the real-space decay is determined by the behavior
of the next term for which $\alpha=\beta+1=3/2$.  To be explicit,
we can define ${\cal A}_n(x)=A_n(x)\,e^{hx}$ so that $\cal A$ is
anti-periodic, ${\cal A}(x+a)=-{\cal A}(x)$, and for large $x$ we find
\begin{eqnarray}
w(x)&\simeq& B_{-1/4}\,{\cal A}_0(x)\,x^{-3/4} \, e^{-hx} \;\;,\nonumber \\
v(x)&\simeq& B_{-1/2}\,\eta_0\,{\cal A}_0(x)\,x^{-1/2} \, e^{-hx} \;\;, \nonumber \\
y(x)&\simeq& (B_{ 1/2}/\eta_0)\,\widetilde{\cal A}_1(x)\,x^{-3/2} \, e^{-hx}
\;\;.
\label{eq:wandecay}
\end{eqnarray}

The above conclusions regarding $y(x)$ rely on the absence of zeros of
$\eta(k)$ inside the strip $-h < {\rm Im}(k) < h$.
If such zeros exist, $y_k(x)=\psi_k(x)/\eta_k$ may have new singularities
and $y(x)$ will then have poor decay compared
to other NWFs. We find that this problem does not arise when using
$t(x)=\delta(x)$ or a narrow Gaussian, but can be triggered by use
of a too-wide Gaussian for $t$.

In view of $n(x',x)=\sum_i w_i(x')\,w_i(x)$ it may appear surprising that
$n(x',x)$ decays more slowly than $w(x)$ ($\alpha_n$=1/2 vs.\ $\alpha_w$=3/4).
The representation of $n$ via $w$ thus has some advantages.  Better yet,
perhaps, one can represent $n(x',x)=\sum_{ij}S_{ij}\,y_i(x')\,y_j(x)$.
Here the slow decay has been transferred to a simple matrix
quantity ($\alpha_S$=1/2) but the NWFs decay very quickly ($\alpha_y$=3/2).

Is it possible to find a NWF with an even faster
decay than $x^{-3/2}\,e^{-hx}$?  Yes; define
a new NWF $z_k=f(k)\,y_k$ where $f(k)$
is analytic in the strip $|{\rm Im}(k)|<h$ and has
simple zeros at the branch points $(2n+1)\pi\pm ih$.  The function
$f(k)=1+\cos(ka)/\cosh(ha)$ is a good candidate \cite{explan-S}.
Then the
leading singularity of $z_k$ is as $(k-k_0)^{3/2}$, and we expect
$z(x)\sim x^{-5/2}e^{-hx}$.  We have confirmed numerically that this
works.  However, since the multiplication by $f(k)$ in $k$-space
corresponds to a convolution in real space, the resulting
$z(x)$ is actually {\it broader} than $y(x)$ or $w(x)$ by almost any
other measure (e.g., second moments \cite{marz-wann}).  Thus, this
strategy may be counterproductive in practice.

Before leaving the 1D case, we make two brief comments.
First, the extension to the case of non-centrosymmetric
potentials in 1D is not difficult, and the results (including values
of the $\alpha$ exponents) are unchanged.
Second, there is an apparent paradox concerning the nearly-free
electron limit.  For free electrons the occupied portion of the
band gives $w(x)\sim\sin(k_Fx)/k_Fx$, i.e., $\sim x^{-1}$.
One may expect this to go over to $\sim x^{-1}\,e^{-hx}$
in the nearly-free case, but this would be
inconsistent with our general result $\alpha=3/4$.
Actually we find there is a crossover behavior, as shown in
Fig.~\ref{fig:power}(b), with $\alpha$=1 behavior for $x<<x_c$
and $\alpha$=3/4 in the true large-$x$ tail.  The crossover distance
$x_c$ increases as the gap decreases; as the gap closes,
$x_c\rightarrow\infty$ and $h\rightarrow 0$.

Before concluding, we briefly discuss the 3D case.  Here the
6-dimensional space of complex $(k_x,k_y,k_z)$
makes the formal analysis difficult.
We have carried out a numerical
calculation of WFs and NWFs for Si using an empirical-pseudopotential
scheme starting from four bond-centered trial functions.  The results are
plotted in Fig.~\ref{fig:lwannier}(b).  Plots of $hx+\ln|F(x)|$
vs. $\ln(x)$ (not shown) again show linear behavior, with slopes
that appear consistent with the 1D values of $\alpha=3/4$, $1/2$, and
$3/2$ for $w$, $v$, and $y$, respectively.  However, in this case we
cannot afford to go to very large $x$ values, and we suspect that there
may be a crossover to larger $\alpha$ values in the far tails.
We leave this as a question for future investigations.

To conclude, we find that in 1D the asymptotic behavior of WFs and
related quantities can all be expressed as $x^{-\alpha}e^{-hx}$
with a common $h$, and with exponents $\alpha$ that take on universal
rational values depending on the type of singularity of the
relevant function at the branch points in the complex-$k$ plane.
It is surprising that this behavior has gone unnoticed since
Kohn's seminal 1959 paper.  The consequences for linear-scaling
calculations, and localized real-space representations of electron
structure more generally, remain to be fully explored.

This work was supported by NSF grant DMR-9981193.




\begin{references}
\bibitem{explan-S}
Alternatively, the overlap $S(k)$ itself can be used
for this purpose via $f(k)=S(k)^{-2}$.  This may be useful since one
may not know the branch-point locations in advance.
\end{references}
\end{document}